\documentclass[aps,pra,twocolumn,amsmath,amssymb,superscriptaddress,floatfix,nofootinbib]{revtex4-1}
\usepackage{graphicx}
\usepackage{multirow}
\usepackage{color}
\usepackage{bm,amsmath}
\usepackage{here}
\usepackage{hyperref}

\def\n{\nonumber \\ }

\usepackage[whole]{bxcjkjatype} 
\begin{document}

\title{Optical response of the Leggett mode in multiband superconductors in the linear response regime}

\author{Takumi Kamatani}
\affiliation{Department of Applied Physics, The University of Tokyo, Hongo, Tokyo, 113-8656, Japan}
\author{Sota Kitamura}
\affiliation{Department of Applied Physics, The University of Tokyo, Hongo, Tokyo, 113-8656, Japan}
\author{Naoto Tsuji}
\affiliation{Department of Physics, The University of Tokyo, Hongo, Tokyo, 113-8656, Japan}
\author{Ryo Shimano}
\affiliation{Department of Physics, The University of Tokyo, Hongo, Tokyo, 113-8656, Japan}
\author{Takahiro Morimoto}
\affiliation{Department of Applied Physics, The University of Tokyo, Hongo, Tokyo, 113-8656, Japan}
\affiliation{JST, PRESTO, Kawaguchi, Saitama, 332-0012, Japan}

\begin{abstract}
We study optical responses of Leggett modes in multiband superconductors in the linear response regime. The Leggett mode is a collective mode unique to multiband superconductors that arises from relative phase fluctuations of superconducting orders for different bands. We use the Ginzburg-Landau (GL) description to study the collective modes in multiband systems. We find that multiband superconductors generally allow a linear coupling between the Leggett mode and external electric fields due to the presence of a cross term between different components of superconducting orders in the GL theory. The presence of a linear coupling for the Leggett mode is in sharp contrast with the absence of that for Higgs (amplitude) modes in single-band superconductors that only support nonlinear optical responses such as third harmonic generation (THG).  We further confirm such a linear coupling in multiband superconductors by a more microscopic description based on a diagrammatic approach. We study the collective modes within the random phase approximation (RPA) and compute their contribution to the linear optical conductivity. These findings suggest a new route to observe the Leggett mode by optical absorption.

\end{abstract}

\date{\today}

\maketitle

\begin{figure*}[t]
  \centering
  \includegraphics[width=\linewidth]{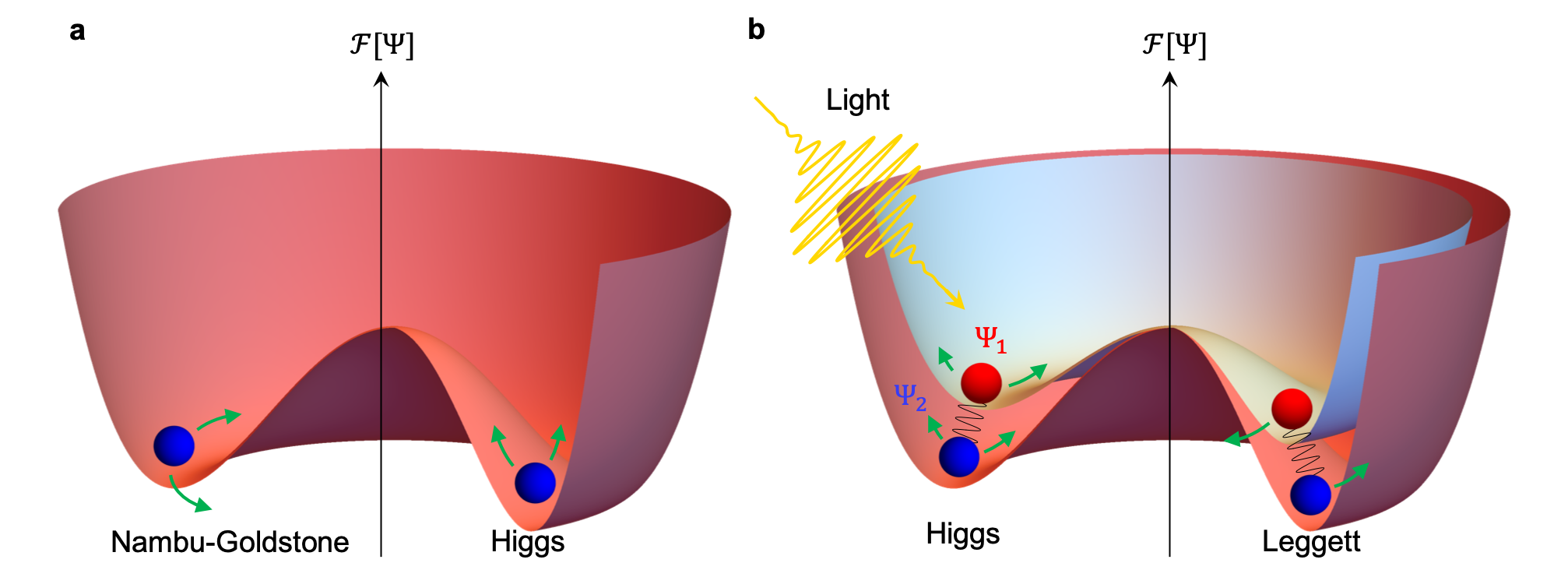}
  \caption{The schematic picture of the free energy of superconductors for (a) a single-band case and (b) a multiband (2 bands) case.
  Higgs modes are the amplitude modes of the order parameter. In multiband superconductors, Leggett mode associated with the fluctuation of the relative phase of two order parameters may appear in the low-energy region, while the overall phase mode (Nambu-Goldstone mode) has an energy scale as large as the plasma frequency due to the Anderson-Higgs mechanism. 
  Light irradiation leads to optical excitation of those collective modes.}
  \label{Fig: collective modes}
\end{figure*}

\section{Introduction}
Superconductivity is one of the most significant phenomena in condensed matter physics \cite{tinkham-textbook}. A macroscopic wave function formed by Copper pair condensation exhibits spontaneous symmetry breaking of the U(1) phase, which is a condensed matter analog of symmetry breaking and mass generation in the Higgs mechanism in high-energy physics \cite{nambu11}. Accordingly, dynamics of the superconducting order parameter $\Psi$ exhibits various collective modes, which is attracting much attention recently \cite{Shimano20}. 

A single-band superconductor supports two collective modes: Nambu-Goldstone (NG) mode and Higgs mode. The NG mode is a collective mode arising from the fluctuation of an overall phase of $\Psi$. The coupling of the NG mode to gauge fields in superconductors lifts its excitation energy to the scale of plasma frequency (which is much higher than the superconducting gap) due to the Anderson-Higgs mechanism \cite{Anderson58,Anderson63,Higgs64,bogoljubov1958new,vaks1962collective}. This situation contrasts with low-energy massless NG modes in charge neutral systems, such as acoustic phonons and magnons in antiferromagnets.  The Higgs mode is a collective mode corresponding to amplitude fluctuation of $\Psi$ and has been named from its analogy to the Higgs particle in elementary particle physics. Higgs mode excitations appear at the energy of the superconducting gap, and have been experimentally observed by Raman spectroscopy in a superconductor coexisting with a charge density wave order \cite{Sooryakumar80,Measson14}, and by terahertz (THz) pump-probe spectroscopy \cite{Matsunaga13,Katsumi18}, and third harmonic generation (THG) \cite{Matsunaga2014,Chu20} in s-wave and d-wave superconductors.

Multiband superconductors host a wider variety of collective modes since more than one superconducting order parameters appear corresponding to multiple band degrees of freedom. For example, multiple Higgs modes appear from the amplitude fluctuation for each band. For the phase fluctuation, one can consider the relative phase fluctuation between different bands, in addition to the overall phase fluctuation. 
Such a relative phase mode is called the Leggett mode~\cite{Leggett66}.
The Leggett mode shows a unique dynamics of the order parameter that reflects pairing symmetry of the multiband superconductors, and its consequence to the Raman signal has been actively explored~\cite{Balatsky00,Burnell10,Ota11,Marciani13,Maiti15,Bittner15,Cea16b,Maiti17}. In particular, observation of the Leggett mode has been reported by Raman spectroscopy for a typical multiband superconductor, MgB$_2$~\cite{Blumberg07}.
The Leggett mode appears around (or below) the energy scale of the superconducting gap as it can avoid the Anderson-Higgs mechanism in contrast to the NG mode.

Studies on optical responses of superconducting collective modes have mostly focused on the nonlinear regime so far \cite{Tsuji15,Kemper15,Krull16,Cea16,Cea16b,Murotani17,Murotani19,Silaev19,Schwarz20,Haenel21}. Indeed, for single-band superconductors, a single photon process for the Higgs mode is prohibited by the inherent particle-hole symmetry \cite{Varma02,Pekker15,Tsuji15}. Instead, a two photon process generates the Higgs mode excitation at the lowest order, as can be understood within both the standard Ginzburg-Landau (GL) description and Anderson pseudospin representation of collective modes \cite{Anderson58,Tsuji15}. In contrast, in multiband superconductors, such a constraint does not apply due to the multiband nature of the order parameter. Hence, a linear coupling between the collective modes and external electromagnetic fields is allowed in principle. Previously, Bogoliubov quasiparticle excitations were shown to support linear optical responses for noncentrosymmetric superconductors \cite{Xu19}. Still, the possibility of linear responses of superconducting collective modes has not been investigated so far.  In particular, it is highly desirable for experiments if the Leggett mode can be detected through optical responses in the linear response regime.

In this paper, we study linear optical responses of collective modes, especially the Leggett mode in multiband superconductors. Using the GL description for multiband systems, we show that a linear coupling term between collective modes and electromagnetic fields is generally allowed from the cross term between the order parameters for different bands. Such linear coupling can be interpreted as an effective Josephson coupling between Cooper pairs of different bands (or orbitals). This picture is further confirmed by a more microscopic description based on a diagrammatic approach. Namely, we study collective excitations within random phase approximation (RPA) and compute diagrams responsible for the linear optical conductivity from the collective modes. We find that the Leggett modes generally appear in the linear conductivity for multiband systems, which contrasts with single-band systems where Higgs mode excitations only appear in nonlinear responses such as THG. Since the Leggett mode may appear below the superconducting gap (in contrast to the Higgs modes that appear right at the superconducting gap), the Leggett mode is shown to form a single peak below the gap in absorption spectra in those situations, which would facilitate experimental observations.

The rest of the paper is organized as follows. In Sec. II, we present the GL description for optical responses in multiband superconductors. Sec. III describes our formalism based on a diagrammatic approach for collective modes and their optical responses. In Sec. IV, we apply our formalism to a simple two-band superconductor and show that the Leggett mode appears in the linear optical conductivity. In Sec. V, we give discussions on experimental feasibility.

\section{Collective modes \label{sec: collective}}
In this section, we give a phenomenological description of collective modes in superconductors in terms of the Ginzburg–Landau (GL) free energy (Fig.~\ref{Fig: collective modes}).
The GL free energy allows us to deduce a coupling between the collective modes and an external light field.

\subsection{Single-band superconductor}

First, let us briefly review the GL description for a single-band superconductor~\cite{Shimano20}. 
We can write down the GL free energy density $\mathcal{F}$ in terms of the superconducting order parameter $\Psi$ as \cite{BauerSigrist,Tsuji15,Shimano20}
  \begin{align}
    \mathcal{F} &=a|\Psi|^2+\frac{b}{2}|\Psi|^4+ \frac{1}{2 m^{*}} |\bm{D}\Psi|^{2}.
    \label{eq: F single}
  \end{align}
Here, $a=a_0(T-T_c)$, $a_0$ and $b$ are some positive constants, $m^*$ is the effective mass of a Cooper pair, $\bm{D}=-i\hbar\bm{\nabla}-e^*\bm{A}$ with the vector potential $\bm A$ and the electric charge of a Cooper pair $e^*=2e$. 
In order to study the coupling between the collective mode and the external electric field, we write the order parameter as
\begin{align}
  \Psi(\bm{r},t)=(\Psi_{0}+H(\bm{r},t))e^{i\theta(\bm{r},t)}.
\end{align}
where $\Psi_{0}=\sqrt{-a/b}$ is the ground state expectation value of the order parameter, $H(\bm{r},t)$ is an amplitude fluctuaion, and $\theta(\bm{r},t)$ is a phase fluctuation.
The terms $H$ and $\theta$ correspond to the Higgs mode and the Nambu-Goldstone mode, respectively, as illustrated in Fig.~\ref{Fig: collective modes}(a).
Due to the Anderson-Higgs mechanism,  the Nambu-Goldstone mode $\theta$ is absorbed into the longitudinal component of the electromagnetic field, and can be eliminated from the low-energy theory by taking the unitary gauge \cite{Anderson63}. 
Then, we obtain the effective Lagrangian for the collective mode (i.e., the Higgs mode $H$), up to a constant term, as
  \begin{align}
    \mathcal{F}&=-2a H^{2} +\frac{\hbar^2}{2 m^{*}} (\bm{\nabla} H)^2
    +\frac{e^{*2}\Psi_{0}^{2}}{2 m^{*}}\bm{A}^2
    \n
    &\qquad 
    + \frac{e^{*2}\Psi_{0}}{m^{*}} \bm{A}^2 H
    +\frac{e^{*2}}{2m^{*}} \bm{A}^2 H^2.
    \label{eq:sin_lag}
  \end{align}
The first and second terms give the mass and the dispersion of the Higgs mode, respectively.
The third term is the mass term for the electromagnetic field.
The last two terms describe nonlinear couplings between the Higgs mode and the external electric field in the second order of $A$.
We find that there is no linear coupling in $A$, which is a direct consequence of the inherent particle-hole symmetry of the superconducting system \cite{Tsuji15}, as we explain in Appendix~\ref{app: microscopic} in detail. 
The current density is obtained from the free energy density as $\bm{j}=-\partial\mathcal{F}/\partial \bm{A}$, 
which means that the current involving the Higgs mode in the leading order is given by
\begin{align}
  \bm{j} \simeq -\frac{2e^{*2}\Psi_{0}}{m^{*}} \bm{A} H.
  \label{eq:current}
\end{align}
The nonlinear coupling $\bm{A}^2 H$ in $\mathcal F$ and the term $\bm{A}H$ in $\bm j$ indicate that the Higgs mode appears in the third-order optical response at the lowest order, for a single-band superconductor.
Indeed the Higgs mode of a superconductor has been observed in third harmonic generation (THG) in the THz regime \cite{Matsunaga2014}.

\subsection{Two-band superconductor}

Next we consider the GL description of collective modes in a multiband superconductor.
For multiband band superconductors, we can consider amplitude and phase fluctuations for each band. Thus we have more collective modes than in a single-band superconductor. 
Here let us focus on the two-band case for simplicity (see Fig.~\ref{Fig: collective modes}(b)). 
In this case, there are two amplitude (Higgs) modes corresponding to two bands.
Two phase modes can be decomposed into the overall phase mode (Nambu-Goldstone mode) and the relative phase mode called the Leggett mode. The Nambu-Goldstone mode is again absorbed into the gauge field and eliminated from the effective theory. 
In contrast, the relative phase mode (the Leggett mode) remains as a physical mode in the theory and can appear in the energy scale of the superconducting gap.
To see this, we consider a GL free energy density for the two-band case given by 
\begin{align}
    \mathcal{F}&=\sum_{i=1,2}\left[a_{i}|\Psi_{i}|^{2}+\frac{b_{i}}{2}|\Psi_{i}|^{4}+
    \frac{1}{2m^{*}_{i}}|\bm{D}\Psi_{i}|^{2} \right]
    \nonumber \\ 
    &+[\epsilon\Psi_{1}\Psi_{2}^{*}
    +\epsilon_{1}(\bm{D}^*\Psi_{1}^{*})\cdot(\bm{D}\Psi_{2})+ c.c.] \nonumber \\
    &+[\bm{d}\cdot(\Psi_{1}^{*}\bm{D}\Psi_{2})+ c.c. ].
    \label{eq: L two band}
\end{align}
Here, in addition to the free energy density for each band with coefficients $a_i,b_i$ and $m_i^\ast$ that are similarly defined as in Eq.~\eqref{eq: F single}, we introduce coupling terms between the two order parameters, $\Psi_1$ and $\Psi_2$, with coefficients $\epsilon, \epsilon_1$ and $d$. 
These coupling terms can be derived from a microscopic treatment as shown in Appendix \ref{app: microscopic}.
While the $\epsilon$ and $\epsilon_1$ terms have been previously incorporated into the GL free energy to study multiband superconductors~\cite{Askerzade,Yerin07,Gurevich03},
the last $d$ term (the linear term in $\bm D$) has not been fully considered in studies of optical responses of the collective modes, to the best of our knowledge.
The presence of the $d$ term can be understood as an effective Josephson coupling between the two states (corresponding to the two bands) which allows a linear coupling to an external electric field.
In fact, this $d$ term plays a central role in linear optical responses of the collective modes, as we explain in the following.

To describe collective modes, we consider fluctuations of the order parameters for the two bands given by
\begin{align}
  \Psi_{1}(\bm{r},t)&=(\Psi_{1,0}+H_{1}(\bm{r},t))e^{i\theta(\bm{r},t)}, \\
  \Psi_{2}(\bm{r},t)&=(\Psi_{2,0}+H_{2}(\bm{r},t))e^{-i\theta(\bm{r},t)},
\end{align}
where $\Psi_{1,0}$ and $\Psi_{2,0}$ are the ground state expectation values of the  order parameters for the two bands, and $H_1$ and $H_2$ describe amplitude fluctuations (Higgs modes) for the two bands. 
Due to the Anderson-Higgs mechanism,
we ignore the overall phase fluctuation and focus on the relative phase fluctuation $\theta(\bm{r},t)$ between the two bands which accounts for the Leggett mode.

Now we expand the effective Lagrangian with respect to $A$ and focus on the coupling term $\mathcal{F}_{EM}$ between the collective modes and the gauge field $\bm{A}$ that is in the linear order of $H_i$ or $\theta$.
The expression for $\mathcal{F}_{EM}$ reads
\begin{align}
    \mathcal{F}_{EM}&=\mathcal{F}_{A^2}+\mathcal{F}_{A1}+\mathcal{F}_{A2}+O(H_i^2,\theta^2, H_i\theta),
\end{align}
with
\begin{align}
    \mathcal{F}_{A^2}&=
    e^{*2}\left(\frac{\Psi_{1,0}}{m_1^{*}}+2\epsilon_1\Psi_{2,0}\right)\bm{A}^2 H_1
    \n &
    +
    e^{*2}\left(\frac{\Psi_{2,0}}{m_2^{*}}+2\epsilon_1\Psi_{1,0}\right)\bm{A}^2 H_2,
\end{align}
\begin{align}
    \mathcal{F}_{A1}&= -e^{*}\hbar\left(\frac{\Psi_{1,0}^2}{m_1^{*}}+\frac{\Psi_{2,0}^2}{m_2^{*}}\right) \bm{A} \cdot \bm{\nabla}\theta,
\end{align}
and
\begin{align}
    \mathcal{F}_{A2}&=-e^{*}\bm{d}\cdot \bm{A}
    (\Psi_{2,0}H_1+\Psi_{1,0}H_2
   -2i\Psi_{1,0}\Psi_{2,0}\theta)+ c.c. 
   \n 
   &=-2e^*\mathrm{Re}[\bm{d}]\cdot \bm{A} (\Psi_{2,0}H_1+\Psi_{1,0}H_2) 
   \n &\quad
   -4e^*\mathrm{Im}[\bm{d}]\cdot \bm{A} \Psi_{1,0}\Psi_{2,0}\theta.
   \label{eq: FA2}
\end{align}
The first term $\mathcal{F}_{A^2}$ describes a nonlinear coupling between the Higgs modes and the electric field for each band, in a similar manner as in the single-band case.
This coupling again results in the THG response of the Higgs modes in multiband superconductors.
The second term $\mathcal{F}_{A1}$ contains a linear term in $\bm A$ with a spatial gradient of the collective mode $\theta$. 
This term may seem to induce a linear coupling between $\bm A$ and the Leggett mode. However, $\bm{A}\cdot \bm{\nabla}\theta$ turns into $\bm{A}\cdot \bm{q}\theta$ with the momentum $\bm q$ upon Fourier transformation, and hence, does not give an effective linear coupling between $\bm A$ and the uniform ($\bm{q}=0$) component of the collective mode that is relevant for optical excitations. 
The last term $\mathcal{F}_{A2}$ is the most important contribution when we consider the linear optical response of collective modes.
Namely, $\mathcal{F}_{A2}$ gives rise to a linear coupling between the collective modes and the electric field through $\bm A H$ and $\bm A \theta$ terms. 
According to Eq.~\eqref{eq: FA2}, the linear coupling to the Higgs modes requires a nonzero real part of $d$, and that to the Leggett mode requires a nonzero imaginary part of $d$.

The presence/absence of a linear coupling between collective modes and an external light field is governed by the inherent particle-hole (PH) symmetry of a superconductor.
In the case of a (conventional) single-band superconductor, it is known that such a linear coupling is absent due to the PH symmetry~\cite{Tsuji15}.
This is the reason why optical responses of collective modes have been mainly studied for nonlinear optical responses such as THG so far~\cite{Murotani17,Shimano20}.
In a stark contrast, the linear coupling term (such as the $d$ term in Eq.~\eqref{eq: L two band}) is generally allowed in multiband superconductors as we show in Appendix \ref{app: microscopic}, even under the constraint of the PH symmetry. 
Namely, the multiband nature allows nonvanishing interband matrix elements of the linear coupling between collective modes and an external light field, enabling an optical excitation of the collective mode in the linear response regime.
In the following, we demonstrate the linear optical response of collective modes in superconductors with a more microscopic calculation based on a diagrammatic approach.

\section{Formalism \label{sec: formalism}}
In this section, we present a microscopic formulation of the electromagnetic response of multiband superconductors.
We describe the collective modes and their optical responses with a diagrammatic approach.

\subsection{Bogoliubov-de Gennes Hamiltonian}
We consider a Hamiltonian of the following form
\begin{equation}
  \mathcal{H}=\mathcal{H}_{0}+ \mathcal{H}_{int},
\end{equation}
where
\begin{align}
  \mathcal{H}_{0}=\sum_{k,\sigma,\alpha,\beta}\xi_{\alpha,\beta}(k) c_{k,\alpha,\sigma}^{\dagger}c_{k,\beta,\sigma}.
\end{align}
is the non-interacting term while $\mathcal{H}_{int}$ represents the (contact) attractive interaction 
\begin{equation}
  \mathcal{H}_{int}=-\frac{1}{N}\sum_{k,k',k'',\alpha}U c_{k'',\alpha,\downarrow}^{\dagger}c_{k,\alpha,\uparrow}^{\dagger}c_{k',\alpha,\uparrow}c_{k+k''-k',\alpha,\downarrow}
\end{equation}
with $U>0$, which is assumed to be diagonal in $\alpha$.
Here $c_{k,\alpha,\sigma}$ annihilates an electron with momentum $k$ and spin $\sigma$, and $\alpha$ represents other internal degrees of freedom (such as a sublattice degree of freedom) which runs from 1 to $n$.
$N=V/V_{unit}$ is the number of unit cells where $V$ is the volume of the system and $V_{unit}$ is the unit cell volume.

To handle the $s$-wave superconductivity, we assume that 
$\langle c_{k,\alpha,\uparrow}c_{k',\alpha,\downarrow}\rangle$ may take a nonzero value if $k=-k'$,
and apply the mean-field approximation to the interaction term.
Then the interaction term reads
\begin{align}
    \mathcal{H}_{int}=\frac{1}{N}\sum_{k,\alpha}(\Delta^{\ast}_{\alpha} c_{k,\alpha,\uparrow}c_{-k,\alpha,\downarrow} +\Delta_{\alpha} c_{-k,\alpha,\downarrow}^{\dagger}c_{k,\alpha,\uparrow}^{\dagger}),\label{eq:int}
\end{align}
up to a constant, where 
\begin{align}
  \Delta_{\alpha}&=-\frac{1}{N}\sum_{k}U\langle c_{k,\alpha,\uparrow}c_{-k,\alpha,\downarrow}\rangle
  \label{eq: Delta}
\end{align}
is the superconducting order parameter.

The obtained Hamiltonian can be concisely expressed in the Nambu representation,
\begin{align}
  \Psi_{k}= (
      c_{k,1,\uparrow},
      \dots,
      c_{k,n,\uparrow} ,
      c_{-k,1,\downarrow}^{\dagger} ,
      \dots,
      c_{-k,n,\downarrow}^{\dagger} 
  )^T.
\end{align}
The mean-field Hamiltonian is represented as the Bogoliubov-de Gennes Hamiltonian with the following form,
\begin{align}
\mathcal{H}_{\text{BdG}}&= \sum_{k} \Psi_{k}^{\dagger} H_{\text{BdG}}(k)\Psi_{k}\\
&= \sum_{k} \Psi_{k}^{\dagger}\left(\begin{array}{cc}
\xi(k) & \Delta\\
\Delta^{\dagger} & -\xi^T(-k)\\
\end{array}\right) \Psi_{k},\label{eq:BdG}
\end{align}
where $[\Delta]_{\alpha,\alpha'}=\Delta_\alpha \delta_{\alpha,\alpha'}$.

\begin{figure}[tb]
  \centering
  \includegraphics[width=0.9\linewidth]{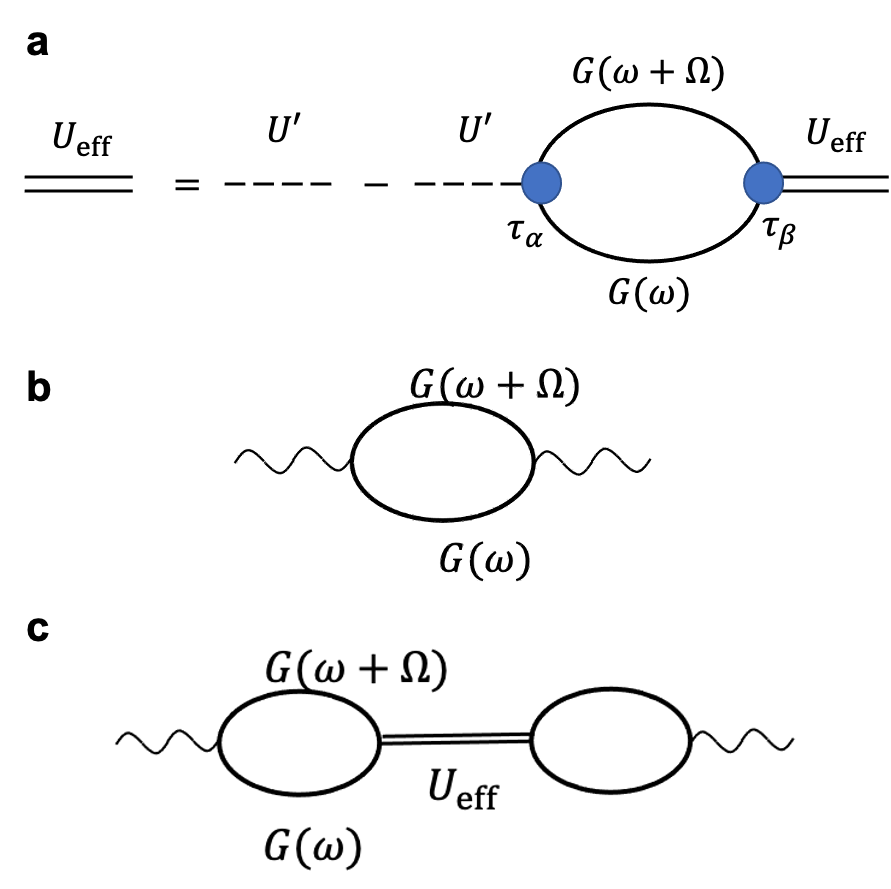}
  \caption{
  Diagrammatic representations for (a) the effective interaction in the random phase approximation and (b,c) the optical conductivity in superconductors. Here
  $U'$ is the bare interaction, while $U_\mathrm{eff}$ is the effective interaction whose poles correspond to collective excitation modes. The optical conductivity is decomposed into (b) Bogoliubov quasiparticles' contribution $\sigma_1(\omega)$ represented by the bubble diagram, and 
  (c) collective modes' contribution $\sigma_2(\omega)$, where poles of the effective interaction $U_\mathrm{eff}$ describe optical absorption from excitations of the collective modes. 
  }
  \label{Figi:dia}
\end{figure}

\subsection{Collective modes and their optical responses}

To formulate the linear response of superconductors, here we introduce the Green function in the Nambu space, $G(i\omega)=(i\omega-H_{\text{BdG}}(k))^{-1}$. 
We can rewrite Eq.~\eqref{eq: Delta} as 
\begin{align}
\Delta_{\alpha} &= U'\int \frac{d \omega}{2 \pi} \int \frac{d k}{(2 \pi)^d}\operatorname{tr}\left[\frac{1}{2} (\tau_{x,\alpha}-i\tau_{y,\alpha}) G(i \omega)\right],
\end{align}
where
\begin{align}
\tau_{x,\alpha}&=
\begin{pmatrix}
O & A_{\alpha}\\
A_{\alpha} &  O
\end{pmatrix}, &
\tau_{y,\alpha}&=
\begin{pmatrix}
O & -iA_{\alpha}\\
iA_{\alpha} &  O
\end{pmatrix}
\end{align}
with $[A_{\alpha}]_{\beta,\beta'}=\delta_{\alpha,\beta}\delta_{\beta,\beta'}$ and $U'=UV_{unit}$.
Without loss of generality, we can make the superconducting gap function real, and replace $\tau_{x,\alpha}-i\tau_{y,\alpha}$ by $\tau_{x,\alpha}$.

Then, we adopt the random phase approximation (RPA) in order to take account of the contribution of the collective modes.
Within the mean-field treatment, the perturbative modulation of the gap function $\Delta_i$ (due to the external field) should also satisfy the self-consistent condition similar to Eq.~\eqref{eq: Delta}. This constraint derived from the self-consistency can be incorporated as an effective interaction, which is diagrammatically represented as the RPA series shown in Fig.~\ref{Figi:dia}(a).
Specifically, the effective interaction is given as 
\begin{align}
    U_\mathrm{eff}&=U'-U'\Pi U'+U'\Pi U'\Pi U'+...\nonumber \\
    &=(1+ U'\Pi)^{-1}U',
    \label{equ:RPA}
\end{align}
where the bubble diagram $\Pi$ is a $2n\times2n$ matrix with indices $(\mu,\alpha)$ ($\mu=x,y$; $\alpha=1,\dots,n$),
whose matrix elements are given as
\begin{align}
&[\Pi(i \Omega)]_{\mu,\alpha,\mu^\prime,\alpha^{\prime}} \nonumber\\&= 
\frac{1}{4}\int \frac{d \omega}{2 \pi} \int \frac{d k}{(2 \pi)^d}\operatorname{tr}[\tau_{\mu,\alpha} G(i \omega+i \Omega) \tau_{\mu^\prime,\alpha^{\prime}} G(i \omega) ] \nonumber\\
&=
\frac{1}{4}\int \frac{d k}{(2 \pi)^d}\sum_{i,j}
\frac{f_{ij}\tau_{\mu,\alpha,ij}\tau_{\mu^\prime,\alpha^{\prime},ji}}{i \Omega-E_{ji}}.
\label{eq:Pi}
\end{align}
Here, $\tau_{\mu,\alpha,ij}\equiv \langle\varphi_{i}|\tau_{\mu,\alpha}| \varphi_{j}\rangle$ with $i,j=1,\dots,2n$ is the matrix element of $\tau_{\mu,\alpha}$ in the band representation, 
i.e., $|\varphi_{i}\rangle$ is the $i$th eigenvector of $H_{\text{BdG}}(k)$ with the eigenvalue $E_i$. 
We have also introduced 
$E_{ij}=E_i - E_j$ and
$f_{ij}=f(E_i) - f(E_j)$ with the Fermi distribution function $f$ at zero temperature, i.e., $f=1$ for occupied bands and $f=0$ for unoccupied bands.
The collective excitation mode appears as a divergence of the effective interaction, i.e., given as a solution of $1+U'\Pi=0$.
We note that the vertices of the bubble diagram $\Pi$ (denoted by the blue circles) can be deduced from Eq.(\ref{eq:int}) by setting $k''=-k$ and rewriting in the Nambu representation as
\begin{align}
    \mathcal{H}_{int}
    &=
     -\frac{1}{N}\sum_{k,k',\alpha}\Psi_{k}^{\dagger}
     \frac{\tau_{x\alpha}+i\tau_{y\alpha}}{2}
     \Psi_{k}
     \Psi_{k'}^{\dagger}
     \frac{\tau_{x\alpha}-i\tau_{y\alpha}}{2}\Psi_{k'}
    \nonumber\\
    &=-\frac{1}{4 N}\sum_{k,k',\alpha}\Big(
    \Psi_{k}^{\dagger}\tau_{x\alpha}\Psi_{k}
    \Psi_{k'}^{\dagger}\tau_{x\alpha}\Psi_{k'}
    \nonumber\\
    &\qquad\qquad\qquad
    +\Psi_{k}^{\dagger}\tau_{y\alpha}\Psi_{k}
    \Psi_{k'}^{\dagger}\tau_{y\alpha}\Psi_{k'}
    \Big).
\end{align}

The linear optical conductivity within the RPA is composed of two diagrams shown in Fig.~\ref{Figi:dia}(b, c). 
The former diagram in Fig.~\ref{Figi:dia}(b) represents the contribution of quasiparticle excitation, and reads
\begin{align}
\sigma_{1}(i \Omega) &= \frac{ie^2}{\hbar^2\omega} \Phi(\omega).
\end{align}
Here $\Phi(\omega)$ is the current-current correlation function and is defined for Matsubara frequency $i\Omega$ as
\begin{align}
\Phi(i \Omega) &= \int \frac{d \omega}{2 \pi} \int \frac{d k}{(2 \pi)^d} \operatorname{tr}[v G(i \omega+i \Omega) v G(i \omega) ] \nonumber \\
&= \int \frac{d k}{(2 \pi)^d} \sum_{i,j}
\frac{f_{ij}v_{ij}v_{ji}}{i \Omega-E_{ji}}.
\end{align}
Here, we used the velocity operator $v$ defined as
\begin{align}
    v(k) &=
    \begin{pmatrix}
    \partial_k \xi(k) &0 \\
    0& -(\partial_k\xi)^T(-k)
    \end{pmatrix}
    =\begin{pmatrix}
    \partial_k \xi(k) &0 \\
    0& \partial_k\xi(k)
    \end{pmatrix}.
\end{align}
and its matrix elements
$v_{ij}\equiv \langle\varphi_{i}|v| \varphi_{j}\rangle$,
where we used the time reversal symmetry $\xi^T(-k)=\xi(k)$ to rewrite the expression for $v$~\cite{Xu19}.
We note that the velocity operator $v$ does not coincide with the $k$ derivative of $H_\mathrm{BdG}$ reflecting the fact that electrons and holes have opposite charges and the gap function does not contribute to the velocity.
Analytic continuation of the Matsubara frequency $i\Omega \to \hbar\omega+i\gamma$ (with $\gamma$ being the level broadening) yields
\begin{align}
    \sigma_{1}(\omega) &=\frac{ie^2}{\hbar^2\omega}\int \frac{d k}{(2 \pi)^d} \sum_{i,j}
\frac{f_{ij}v_{ij}v_{ji}}{\hbar \omega-E_{ji}+i\gamma} \nonumber \\
&=\frac{\pi e^2}{\hbar^2\omega}\int \frac{d k}{(2 \pi)^d} \sum_{i,j}
f_{ij}v_{ij}v_{ji} 
\delta(\hbar\omega-E_{ji}),
\label{eq: sigma1}
\end{align}
where we assumed a sufficiently small $\gamma$ in the last line.

The latter diagram in Fig.~\ref{Figi:dia}(c) corresponds to the linear optical conductivity through the collective excitation mode, which is given as 
\begin{align}
  \sigma_{2}(\omega) = \frac{ie^2}{\hbar^2 \omega}Q^T(\omega) U_\mathrm{eff}(\omega) Q^{\ast}(-\omega),
  \label{eq:ss2}
\end{align}
where $Q(\omega)$ is a vector defined as
\begin{align}
[Q(i\Omega)]_{\mu,\alpha} &=\int \frac{d \omega}{2 \pi} \int \frac{d k}{(2 \pi)^d}\operatorname{tr}[v G(i \omega+i \Omega) \tau_{\mu,\alpha} G(i \omega) ] \nonumber\\
&=\int \frac{d k}{(2 \pi)^d}\sum_{i,j}
\frac{f_{ij}v_{ij} \tau_{\mu,\alpha,ji}}{i \Omega-E_{ji}}.
\label{eq:q}
\end{align}
Again, the expressions for $Q(\omega)$ and $U_\mathrm{eff}(\omega)$ in terms of real frequencies are obtained with an analytic continuation $i\Omega \to \hbar \omega +i\gamma$.
Optical excitations of collective modes arise from the poles in $U_\mathrm{eff}$ which is specified by $\omega$ satisfying $1+U'\Pi(\omega)=0$. This means that $Q(\omega)$ is off resonant, i.e., contributions to $Q(\omega)$ come from the terms with nonzero denominator ($\hbar\omega -E_{ji} \neq 0$).

\section{Application to 1D two-band model\label{sec: application}}
In this section, we investigate the linear response of superconductors using the above formulation.
While conventional single-band superconductors do not support direct (linear) coupling between collective modes and electromagnetic fields, a multiband nature of the gap function  allows an optical excitation of collective modes in the linear response regime.
As a minimal example for multiband superconductors, we adopt a model for two-band superconductors based on the 1D Rice-Mele model \cite{RiceMele}.

\begin{figure*}[tb]
  \centering
  \includegraphics[width=0.8\linewidth]{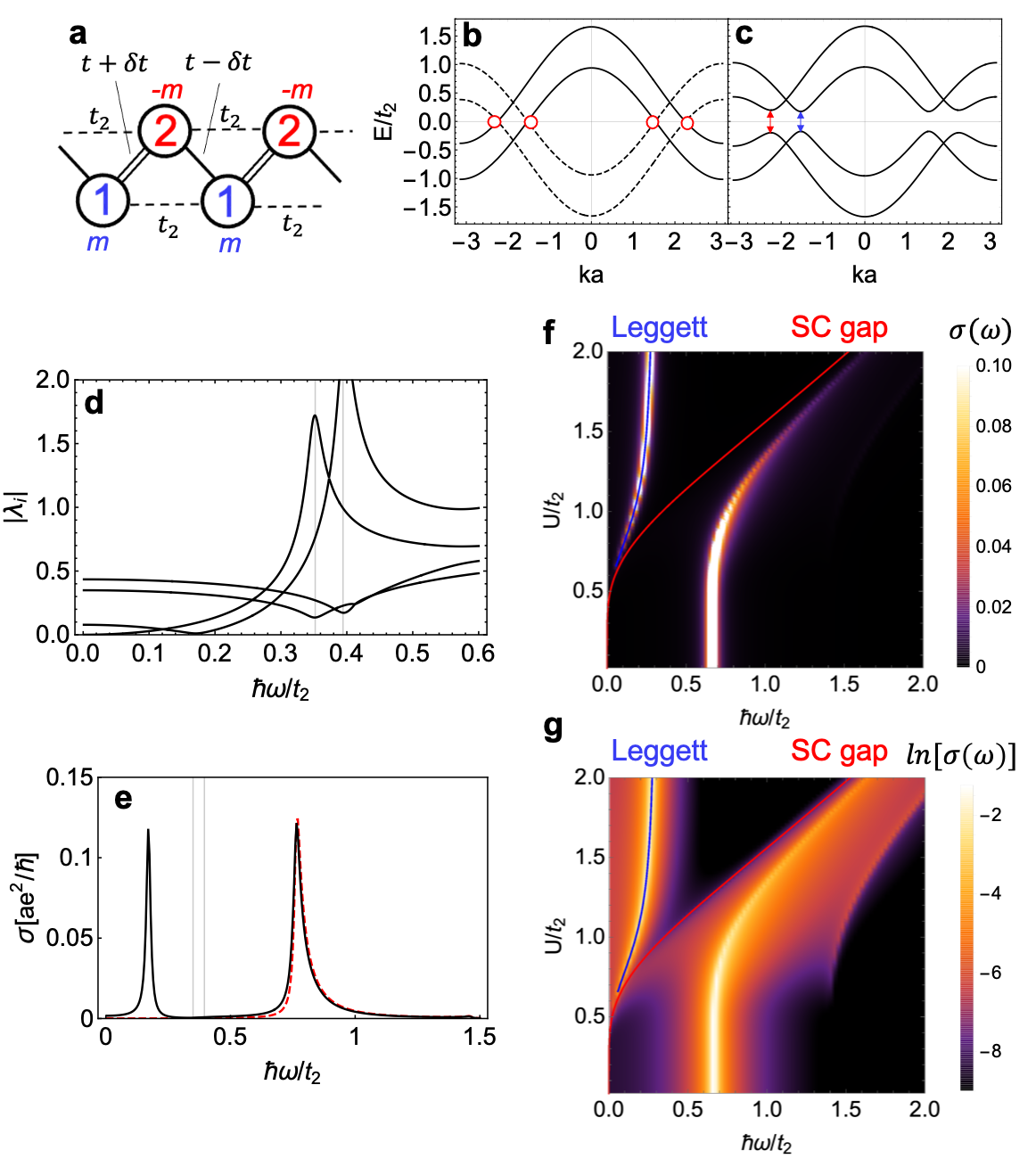}
  \caption{
  Optical response of the collective modes in the 1D model [Eq.\eqref{eq: Rice Mele + U}] for two-band superconductors.
  (a) A schematic picture for the 1D model. We adopt the Rice-Mele model with the next-nearest-neighbor hopping $t_2$ and the attractive interaction $U$. The nearest-neighbor hopping has the uniform part $t$ and the bond alternating part $\delta t$. $m$ is the staggered potential.
  (b, c) Band structures 
  (b) in the normal state with $U=0$, 
  and (c) in the superconducting state with $U=1$. 
  The hole branch is depicted with the dashed lines and the Fermi points are indicated by the red circles in (b). 
  Two superconducting gaps are indicated by the red and blue arrows in (c).
  (d) Absolute values $|\lambda_i|$ of the eigenvalues $\lambda_i$ of $1+U'\Pi$ for $U=1$. Zero eigenvalues ($\lambda_i=0$) correspond to the superconducting collective modes.
  The zero eigenvalues at $\hbar\omega=0$ and $\hbar\omega\simeq0.2t_2$ correspond to the Nambu-Goldstone mode and the Leggett mode, respectively. 
  The two cusps appearing at $\hbar\omega=2\Delta_L, 2\Delta_U$ correspond to the two Higgs modes.
  The gray vertical lines represent energies of the two superconducting gaps.
  (e) Optical conductivity $\sigma(\omega)$ as a function of the frequency for $U=1$. The red dashed line represents a contribution from Bogoliubov quasiparticles. The solid line represents the sum of the contributions from Bogoliubov quasiparticles and the collective modes. The vertical gray lines show the two superconducting gaps where the Higgs modes appear. Below the superconducting gaps (the gray lines), a large peak appears from the Leggett mode excitation, clearly indicating that the Leggett mode is accessible by optical absorption.
  (f, g) Interaction dependence of the optical conductivity $\sigma(\omega)$.
  $\sigma(\omega)$ is color coded as a function of the frequency $\omega$ and the attractive interaction $U$ (f) in the linear scale, and (g) in the log scale.
  The Leggett mode appears below both the superconducting gap and the band gap in the present model.
  We use the parameters, 
    $t=0.2$, $\delta t=0.1$, $m=0.3$, and $t_{2}=1$.
  }
  \label{Figi:Figi}
\end{figure*}

\subsection{Two-band superconductor model based on the Rice-Mele model}
\begin{figure*}[t]
  \centering
  \includegraphics[width=\linewidth]{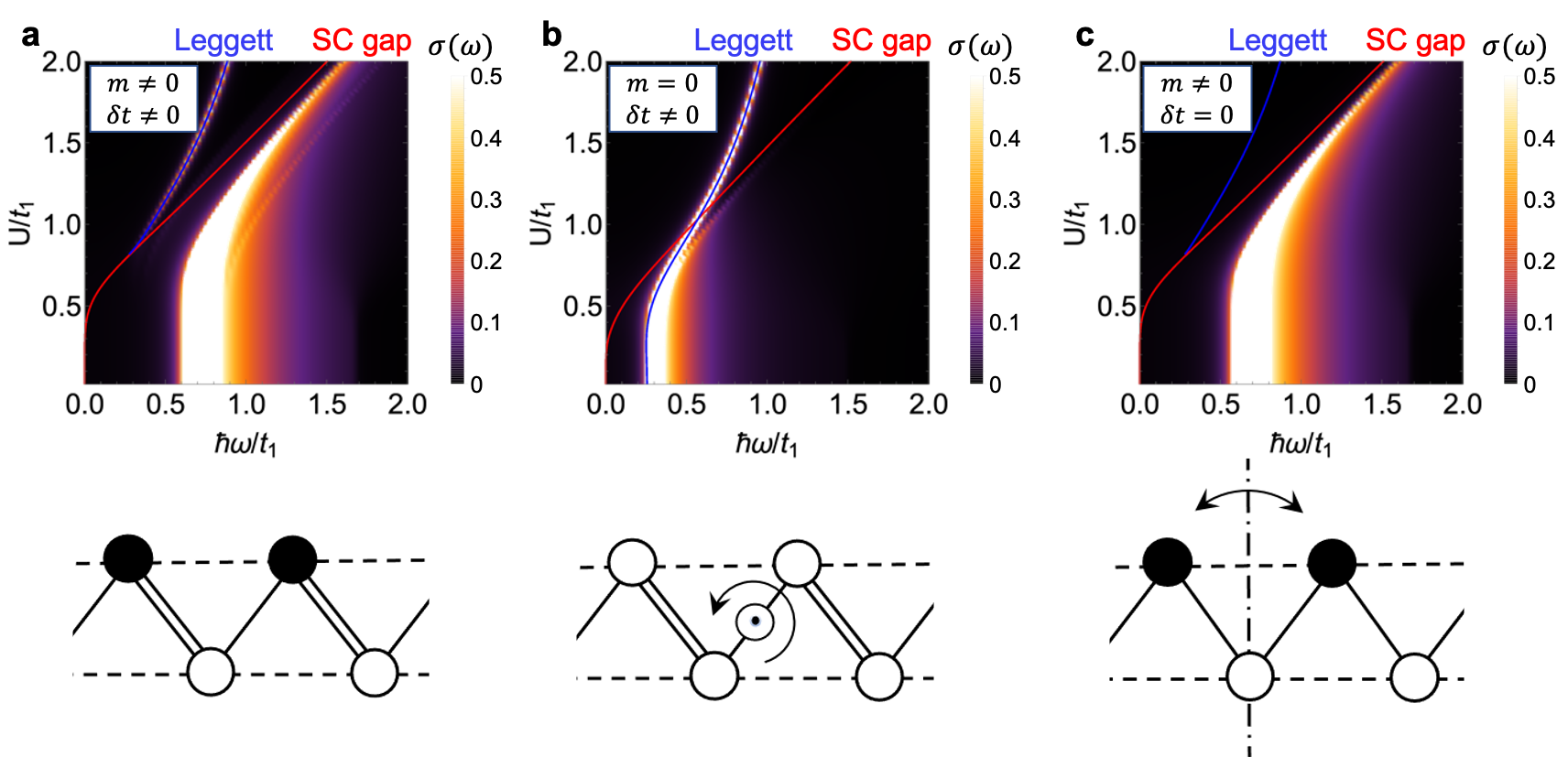}
  \caption{Optical responses  with collective excitations as a function of  interaction and frequency in (a) a system with $m \neq 0$ and $\delta t\neq0$, (b) a system with $m=0$ $\delta t \neq 0$ and (c) a system with $m \neq 0$ $\delta t =0$. The blue line represents the peak positions of Leggett mode and the red line represent superconducting gap energy.
  Lower panels show the schematic pictures of the lattice structure and spatial symmetries: The bond-centered $C_2$ rotation (bond-centered inversion in purely 1D cases) symmetry is recovered for (b), while the site-centered mirror (site-centered inversion) for (c).
    }
  \label{Figi:t_sym}
\end{figure*}

We consider the Rice-Mele model with the next-nearest-neighbor hopping $t_2$ on a 1D chain and an attractive onsite interaction $U$, 
which is depicted in Fig.~\ref{Figi:Figi}(a). The Hamiltonian reads
\begin{align}
    \mathcal{H}&=
    \sum_{i,\sigma}
    \{
    [t+ (-1)^i\delta t]c_{i+1,\sigma}^{\dagger}c_{i,\sigma}
    +t_2 c_{i+2,\sigma}^{\dagger}c_{i,\sigma}
    +h.c. \}
    \nonumber \\
    &+ \sum_{i,\sigma} (-1)^i m c_{i,\sigma}^{\dagger}c_{i,\sigma}
    -\mu \sum_{i,\sigma} c_{i,\sigma}^{\dagger}c_{i,\sigma}
    -U \sum_{i} n_{i,\uparrow}n_{i,\downarrow},
    \label{eq: Rice Mele + U}
\end{align}
where $t+ (-1)^i\delta t$ is the nearest-neighbor hopping with a staggered component, $m$ is the staggered onsite potential, and $\mu$ is the chemical potential. 
The annihilation operator for an electron at the $i$th site with spin $\sigma$ is denoted as $c_{i,\sigma}$, and $n_{i,\sigma} = c_{i,\sigma}^{\dagger}c_{i,\sigma}$.
The Rice-Mele model is known as one of the minimal models for noncentrosymmetric systems, where $m\neq0$ breaks the bond-centered inversion symmetry while $\delta t\neq0$ breaks the site-centered one. These symmetries correspond to the (bond-centered) $C_2$ rotation along the $z$-axis and the (site-centered) mirror along the $yz$ plane, respectively, for the geometry shown in Fig.~\ref{Figi:Figi}(a) [See also the lower panels of Fig.~\ref{Figi:t_sym}].

We define the annihilation operator in the momentum space (introduced in Sec.~\ref{sec: formalism}), $c_{k,\alpha,\sigma}$ with $\alpha=1,2$,
as the Fourier transform within each sublattice, 
\begin{align}
    c_{k,1,\sigma}&=\frac{1}{\sqrt{N}}\sum_j c_{2j,\sigma}e^{-ikja}, \\
    c_{k,2,\sigma}&=\frac{1}{\sqrt{N}}\sum_j c_{2j+1,\sigma}e^{-ik(j+1/2)a}.
\end{align}
Here, $a$ is the lattice constant 
which we set as $a=1$ below for simplicity, and $N$ is the total number of unit cells (i.e.,  the system consists of $2N$ sites).
Then the Bogoliubov-de Gennes Hamiltonian after the mean-field approximation is obtained as
\begin{align}
    \mathcal{H}_{\text{BdG}} 
    &=\sum_{k} \Psi_{k}^{\dagger}\left(\begin{array}{cc}
\xi(k) & \Delta\\
\Delta^{\dagger} & -\xi^{T}(-k)\\
\end{array}\right) \Psi_{k}
\label{eq: RM BdG}
\end{align}
where
\begin{align}
    \xi(k)=\begin{pmatrix}
2t_2 \cos k+m-\mu & 2t\cos \frac{k}{2}+2i\delta t\sin \frac{k}{2}\\
2t\cos \frac{k}{2}-2i\delta t\sin \frac{k}{2} &  2t_2 \cos k-m-\mu
\end{pmatrix},
\label{eq: xi}
\end{align}
and
\begin{align}
\Delta=
\begin{pmatrix}
\Delta_1 & 0\\
0 &  \Delta_2
\end{pmatrix}.
\end{align}
We note that $\mathcal{H}_{\text{BdG}}$ preserves the time-reversal symmetry (TRS) $\mathcal{T}$ which can be represented as $\mathcal{T}=K$ ($K$: complex conjugation).

\subsection{Results of the optical responses from collective modes}
Let us first check that the present model indeed describes a multi-gap superconductor when the model parameters are appropriately chosen.
We set $U=1$, $\mu=0.3$, $t=0.2$, $\delta t=0.1$, $m=0.3$, and $t_{2}=1$. 
Note that $t\pm\delta t<t_{2}$ holds for the present choice, which can be realistic, e.g., when the configuration depicted in Fig.~\ref{Figi:Figi}(a) is realized as a ladder structure of two superconducting chains. 
Due to the above condition, the energy dispersion for the noninteracting case ($U=0$) has two pairs of Fermi points (corresponding to two Fermi surfaces in higher dimensions), as shown in Fig.~\ref{Figi:Figi}(b). This implies that a multi-gap superconductivity appears when the attractive interaction is switched on. 
Indeed, the superconducting ground state with $U=1$ is characterized by two different gap energies, $2\Delta_{L}=0.35t_2$ and $2\Delta_{U}=0.40t_2$,
which can be seen in the energy dispersion of the Bogoliubov quasiparticles in Fig. \ref{Figi:Figi}(c) (blue and red arrows).

Next, let us explore the characteristic frequency for the collective excitation, 
which can be deduced from the condition that one of the eigenvalues $\lambda_i$ of $1+U'\Pi=U'U_\mathrm{eff}^{-1}$ becomes zero [See Eq.~\eqref{equ:RPA}].
We show the frequency dependence of $|\lambda_i|$ (the absolute values of the eigenvalues of $1+U'\Pi$) in Fig. \ref{Figi:Figi}(d).
We can find that $|\lambda_i|$ shows two zeros and two local minima with cusp structures. 
The zero at $\hbar\omega=0$ corresponds to the Nambu-Goldstone mode, which should have no physical effect since it can be eliminated using the gauge transformation. The cusps appearing at $\hbar\omega=2\Delta_L,2\Delta_U$ correspond to the two Higgs modes, whose contributions are expected to be small in the present case since the eigenvalues do not reach zero. The other zero at the frequency $\hbar\omega\simeq0.2t_2$ should be assigned to the Leggett mode. 
While the energy of the Leggett mode varies with parameters in general, we find that the Leggett mode appears below the superconducting gap for the present set of parameters, which facilitates distinguishing the contribution of the Leggett mode to conductivities as we show below.  

Now we investigate the role of the collective excitation modes in the optical response. 
First, we show the optical response of the quasiparticles $\sigma_1$ [See Fig.~\ref{Figi:dia}(b)] 
by the dotted line in Fig.~\ref{Figi:Figi}(e). We can see a prominent peak at $\hbar\omega\simeq0.8t_2$, which is far above the superconducrting gap. This peak originates from the interband transition across the band gap, and thus also appears in the normal state without a superconducting order (for $U=0$).
On the other hand, if we take account of the collective excitation modes via the diagram shown in Fig.~\ref{Figi:dia}(c), 
we obtain the optical response drawn by the solid line in Fig.~\ref{Figi:Figi}(e).
Another peak appears at the frequency characterizing the Leggett mode at $\hbar\omega=0.2t_2$ [See Fig.~\ref{Figi:Figi}(d)].
Namely, this result demonstrates that we can optically excite the Leggett mode in multiband superconductors even within the linear response regime, 
as opposed to single-band superconductors that require nonlinear optical effects to access the collective modes. 
This is the key result of the present paper.

To further clarify the structure of the optical response, we show the interaction dependence of the optical conductivity in Fig.~\ref{Figi:Figi}(f,g), with linear and logarithmic scales.
The red line indicates the superconducting gap, while the blue line shows the peak position of the Leggett mode. 
As we have seen, the peak far above the superconducting gap is characterized by the band gap between two bands in the normal state, and remains nonzero at $U=0$. 
In the logarithmic scale, we can see that the Bogoliubov quasiparticles (appearing at the higher energy region than the red curve) also contribute to the optical conductivity.
We can clearly see that the Leggett mode appears below both the superconducting gap and the band gap in the present model, which is a suitable situation for observing the Leggett mode in the linear optical conductivity.

\subsection{Symmetry consideration \label{sec: symmetry}}
For the nonzero linear response shown above, symmetries of the system as well as the multi-gap nature play the key role.
Here we discuss how the presence of the symmetry restricts the coupling between electromagnetic fields and the collective excitations, which is governed by the vector $Q_{\mu,\alpha}$ defined in Eq.~(\ref{eq:q}).
For instance, as we detail in Appendix~\ref{app: symm}, $Q_{\mu,\alpha}(i\Omega)$ must be real due to the inherent particle-hole symmetry. The main interest here is that what kind of symmetry leads to constraints that allow/prohibit the linear coupling of the collective modes.

Since $Q_{\mu,\alpha}$ is odd in the velocity operator $v$, the linear response vanishes if there is a symmetry that inverts $v$ to $-v$.
In the present model [Eq.~(\ref{eq: xi})], such a cancellation occurs in the system with $\delta t=0$, i.e., the site-centered inversion symmetric case. For the geometry shown in Fig.~\ref{Figi:Figi}(a), this symmetry corresponds to the site-centered mirror, as depicted in the bottom panel of Fig.~\ref{Figi:t_sym}(c).  We can indeed show that $Q_{\mu,\alpha}=-Q_{\mu,\alpha}$ (see Appendix~\ref{app: symm}).

On the other hand, if such a symmetry operation inverting $v$ also transforms the gap function $\Delta_i$, the $Q$ vector may have nonzero elements.
When $m$ is set to zero in the present model, the system has the bond-centered inversion symmetry, which can also be interpreted as the $C_2$ rotational symmetry as depicted in the bottom panel of Fig.~\ref{Figi:t_sym}(b). This symmetry is accompanied by the interchange of $\Delta_1$ and $\Delta_2$ as opposed to the site-centered symmetry.
Thus the constraint on the $Q$ vector due to the symmetry becomes $Q_{\mu,1}=-Q_{\mu,2}$. This implies that $Q$ is orthogonal to the eigenvectors of $U_\text{eff}$ corresponding to the uniform oscillation of the amplitude/phase, but is allowed to have an overlap with the Leggett mode.

In general, the linear coupling between the Leggett mode and the electromagnetic field is forbidden under the inversion symmetry, when the basis that diagonalizes the gap function is invariant under the inversion operation. This can be derived from an action of the inversion operator on the GL free energy as described in Eq.~\eqref{eq: Fij q} in Appendix A. 
In the present case, the gap function is diagonal with respect to the site basis. Thus the site-centered inversion forbids the linear coupling, while the bond-centered inversion does not.

Let us confirm the above observation by numerical calculations. 
We here consider the system with $\mu=0.66$, $t=1$, and $t_{2}=0.8$ (Note that they are different from the previous subsection).
We show the interaction and frequency dependence of the optical conductivity through the collective excitation mode in Fig.~\ref{Figi:t_sym}.

We first compare the result for $(m,\delta t)=(0.3,0.1)$ [Fig.~\ref{Figi:t_sym}(a)] and $(m,\delta t)=(0,0.1)$ [Fig.~\ref{Figi:t_sym}(b)]. 
The blue and red lines in the figure indicate the peak positions of the Leggett and Higgs modes (at the superconducting gap point), respectively, which are deduced from the frequency dependence of the eigenvalues of $U_\text{eff}^{-1}$, as in the previous subsection. 
We can confirm that the peak corresponding to the Leggett mode still exists even if the site-centered inversion symmetry is recovered.
On the other hand, when we set $(m,\delta t)=(0.3,0)$ [Fig. \ref{Figi:t_sym}(c)], the peak due to the Leggett mode dissapears, while the effective interaction $U_{\text{eff}}$ has the pole corresponding to the Leggett mode (drawn by the blue line).

In addition to the Leggett mode, the Higgs modes also exhibit nonzero optical weights when $\delta t\neq 0$.
Specifically, we can numerically confirm the eigenvectors corresponding to the Higgs modes in $U_\mathrm{eff}(\omega)$ and the vertex $Q(\omega)$ have a nonzero overlap. 
Actually, the GL description in Sec.~\ref{sec: collective} forbids such linear coupling between the Higgs modes and external electric fields, since the $d$ term is purely imaginary as a consequence of the particle-hole symmetry as we show in Appendix~\ref{app: microscopic}.
However, the GL free energy is obtained in the static limit. 
In contrast, 
the RPA treatment for the collective modes in this section incorporates the dynamical effects from the finite frequency $\omega$ in $U_\mathrm{eff}$, which allows a nonzero optical weight for the Higgs modes. 
Since such a dynamical effect is relatively small, the optical weights for the Higgs modes are smaller than that for the Leggett mode.

\section{Discussions}
In this paper, we have studied optical responses of multiband superconductors and showed that the Leggett mode is generally accessible by a linear optical probe in multiband superconductors. 
While a linear coupling of the collective (Higgs) mode to the external light field is forbidden in a single-band superconductor due to the inherent PH symmetry, the multiband nature enables the linear coupling between the Leggett mode and the external light field.
This observation is understood both by the GL description and the random phase approximation for the collective modes.
In contrast to the Higgs modes that appear at the superconducting gaps, the Leggett mode can appear either below or above the superconducting gaps. In particular, when the Leggett modes appear below the superconducting gaps, the optical conductivity shows a clear peak structure at the Leggett mode excitation, which would facilitate optical observations of the Leggett mode.

We estimate the optical weight of the Leggett mode peak from the expressions in Eqs.~\eqref{eq: sigma1} and \eqref{eq:ss2}.
We consider that the velocity matrix element $v_{ij}$ has the order of the Fermi velocity $v_F$, the energy difference can be estimated to be the Fermi energy $E_F$, and the matrix elements of Pauli matrices in the Nambu space $\tau_\alpha$ are of the order of 1.
These rough estimations give the optical weights for the interband transition (in the normal state) $W_N$ and for the Leggett mode excitation $W_L$ as
\begin{align}
 W_N&=\int_0^{E_F} d\omega \sigma_1 \simeq (e^2 /\hbar)(v_F^2/a E_F) \\
 W_L&=\int_0^{E_L} d\omega \sigma_2 \simeq (e^2/\hbar)(Ua/\hbar^2)
\end{align}
where 
$E_L$ is the energy of the Leggett mode (of the order of the superconducting gap) and $a$ is the lattice constant.
Thus the ratio of the optical weight for the Leggett mode and the interband transition is given by
\begin{align}
    \frac{W_L}{W_N}&\simeq \frac{U}{E_F},
\end{align}
by assuming $E_F\simeq v_F (\hbar/a)$.
This relationship indicates that the relative optical weight of the Leggett mode is crudely determined by the ratio between the attractive interaction that induces the superconductivity and the Fermi energy, which is roughly deduced from $T_c/T_F$ (where $T_c$ is the superconducting transition temperature and $T_F$ is the Fermi temperature).
Multiband superconductors can generally be considered as candidate materials to observe the Leggett mode through the linear optical repsponse. 

For multiband superconductors such as MgB$_2$~\cite{MgB2Rev},
the optical weight of the Leggett mode is expected to be $W_L/W_N \simeq 10^{-3}$.
For high T$_c$ superconductors such as cuprates with multilayers, we can expect a larger response with $W_L/W_N \simeq 10^{-2}$.
Even larger response can possibly be attained in an iron-based superconductor FeSe, which has been suggested to lie at the BCS-BEC crossover regime \cite{Lubashevsky12,Okazaki14,Kasahara14}, where the superconducting gap size becomes comparable to the Fermi energy. The optical response of the Leggett mode in iron-based superconductors is also interesting since it can elucidate the symmetry of order parameters, including the time-reversal symmetry breaking. Accordingly, the visibility of the Leggett mode in particular in the Raman signal has been intensively investigated~\cite{Chubukov09,Burnell10,Marciani13,Maiti15,Khodas14,Huang18}. 
Two dimensional materials such as transition metal dichalcogenides (TMDs) are also interesting multiband superconductors which may allow observations of the Leggett modes in optical responses~\cite{TMDSC,wan2021observation}.

For applications to real materials, impurity effects have been known to play an important role in observing collective modes of superconductors by means of optical responses \cite{Jujo2018,Murotani19,Silaev19,Shimano20}. In particular, the visibility of the Leggett mode in THG measurements is strongly affected by impurity scattering \cite{Murotani19}. Whether similar effects may be observed in the linear-response regime is an important issue, which is, however, beyond the scope of the present study, and will be left as a future problem. Here we emphasize that in principle the linear coupling between the Leggett mode and light is allowed, apart from quantitative aspects, as shown in the present paper. Another issue is the effect of screening due to long-range Coulomb interactions, which deserves for further considerations. In the case of Higgs modes, the screening effect does not give significant contributions to nonlinear responses \cite{Cea16}. In the case of the Leggett mode, the screening effect has been discussed not to be so relevant since the interband current will flow over wave vectors associated to the size of Wannier orbitals \cite{Murotani17}.

\begin{acknowledgments}
We thank Takashi Oka for fruitful discussions.
This work was supported by 
JST CREST (Grant No. JPMJCR19T3) (S.K., R.S., T.M.),
JST PRESTO (Grant No. JPMJPR19L9) (T.M.),
and KAKENHI (Grant No. JP20K03811) (N.T.).
\end{acknowledgments}

\appendix

\section{Microscopic derivation of the $A$ linear term in the effective Lagrangian
\label{app: microscopic}}

In this section, we present a microscopic derivation of the $A$ linear term in the effective Lagrangian for two-band superconductors.
To this end, we start from the Rice-Mele model with an attractive interaction in Eq.~\eqref{eq: Rice Mele + U} and derive the effective Lagrangian.
For simplicity, we set $t_2=0$ in this section.

For the microscopic Hamiltonian Eq.~\eqref{eq: Rice Mele + U}, a GL free energy density $\mathcal{F}$ for the gap function $\Delta$ is written as~\cite{Wakatsuki17,Wakatsuki18,Hoshino18}
\begin{align}
    \mathcal{F}&=
    \int \frac{dk}{2\pi} \left[
    \sum_{i=1,2} \frac{|\Delta_i|^2}{U}
    +\sum_{i,j=1,2} 
    \Delta_i^* F_{ij}(q) \Delta_j
    \right],
    \label{eq: F appendix}
\end{align}
with
\begin{align}
    F_{ij}(q)&=
    T\sum_{n} \int \frac{dk}{2\pi}g_{ij}(i\omega_n, k+q/2) g_{ij}(-i\omega_n, -k+q/2).
    \label{eq: Fij q}
\end{align}
Here, $i,j$ are site indices within the unit cell, 
and $g$ is the normal part of the Green's function given by
\begin{align}
    g(i\omega_n, k)&=
    \frac{1}{i\omega_n - \xi(k)}.
\end{align}
We took the convention, $k_B=1$ for simplicity.
The gap function is proportional to the superconducting order parameter as $\Delta_i = C \Psi_i$ with a constant $C$, which ensures a direct relationship between Eq.~\eqref{eq: F appendix} and the free energy density discussed in Sec.~\ref{sec: collective}.
The $q$ dependence in $F_{ij}$ describes the dynamics of the order parameter under the external electric field. Namely, the coupling to the gauge potential is given by rewriting $\mathcal{F}$ in the real space representation and introducing a minimal coupling to $A$, which results in substituting $\hbar q \to -i\hbar\nabla +e^*A $ in $\mathcal{F}$.
Therefore, we can obtain the $A$ linear in $\mathcal{F}$ by looking at the $q$ linear term in $F(q)$.

We consider a series expansion of $F(q)$ with respect to $q$ as \begin{align}
    F(q)&= f_0 + f_1 q + f_2 q^2 + O(q^3).
    \label{eq: F q}
\end{align}
Here, $f_0$ gives the quadratic term in the order parameter such as $a_i$ and $\epsilon$ terms in Eq.~\eqref{eq: L two band}, and
$f_2$ gives the kinetic term such as $1/2m^{*}_i$ and $\epsilon_1$ terms in Eq.~\eqref{eq: L two band}.
In addition, $f_1$ gives the $A$ linear term (the $d$ term in Eq.~\eqref{eq: L two band}) that we are interested in.
The $f_1$ term arises from the fact that the base of the gap function $\Delta$ and that of the energy band $\xi$ are different, and an effective pair hopping takes place during the propagation of the Cooper pair, which resembles a Josephson coupling~\cite{Anishchanka07}.   
Now, substituting Eq.~\eqref{eq: xi} in $F(q)$ and collecting the $q$ linear term gives
\begin{align}
    f_1&=
    \begin{pmatrix}
    0 & \tilde d \\
    \tilde{d}^* & 0 \\
    \end{pmatrix}
    +O(\delta t^2),
\end{align}
where $\tilde d$ is defined as
\begin{align}
    \tilde d&= T\sum_n\int\frac{dk}{2\pi}
    \frac{-i \delta t}{[(i\omega_n -\mu)^2-\epsilon(k)^2][(i\omega_n +\mu)^2-\epsilon(k)^2]},
\end{align}
with $\epsilon(k)=\sqrt{4 t^2 \cos^2 k/2 + m^2}$. 
A dominant contribution in the Matsubara frequency ($i\omega_n$) summation comes from small $n$. Focusing on the contributions from small $n$ and performing the $k$ integration, we obtain an expression for the linear coupling term $d=(|C|^2/\hbar) \tilde d$ in Eq.~\eqref{eq: L two band} as
\begin{align}
    d 
    &\simeq \frac{|C|^2}{\hbar} T\sum_n \frac{-i\delta t}{4\mu^2 v_F \omega_n} \n 
    &\simeq -i \frac{|C|^2}{\hbar} \frac{\delta t}{8\pi\mu^2 v_F} S_1(T).
    \label{eq: d microscopic}
\end{align}
Here we used $S_1(T)=\ln(2e^\gamma E_c/\pi T)$ where $\gamma$ is the Euler's constant and $E_c$ is the energy cutoff \cite{BauerSigrist,Hoshino18}. 
We note that this effective Lagrangian is valid for near superconducting transition temperature ($T_c$), and a different vortex dynamics plays an important role in lower temperatures, especially below the Kosterlitz-Thouless transition temperature ($T_{KT}$).
Thus the logarithmic divergence with $T$ in Eq.~\eqref{eq: d microscopic} is cutoff at the energy scale of $T_c$ \cite{Wakatsuki17,Hoshino18}.

The derived $q$ linear term in $F$ indicates the coupling between the collective modes and external light field in the linear regime.
The nonzero $d$ term in Eq.~\eqref{eq: d microscopic} is purely imaginary and results in linear coupling between the Leggett mode and an external light field as we discussed in Sec.~\ref{sec: collective}. Indeed this is what we observed from a diagrammatic derivation of the linear conductivity in Sec.~\ref{sec: application} for Rice-Mele model with an attractive interaction.
The appearance of $\delta t$ in the expression for $d$ means that breaking of bond center inversion is necessary for the linear coupling between the Leggett mode and the external light, which is consistent with the results in Sec.~\ref{sec: symmetry}. 

The inherent particle-hole (PH) symmetry gives a constraint on the form of the linear coupling term $f_1$ as follows.
Let us consider the particle-hole transformation for the integrand in the expression for $F_{ij}$ in Eq.~\eqref{eq: Fij q}.
The (inherent) particle-hole symmetry is given by exchanging the positive energy mode with the negative energy mode of its time-reversal partner. Specifically, the particle-hole transformation results in 
\begin{align}
    &\mathrm{PH}:& 
    i\omega_n &\to -i\omega_n, 
    &
    \xi(k) &\to U_T \xi(-k)^T U_T^{-1},
\end{align}
in Eq.~\eqref{eq: Fij q}, where $U_T$ is a unitary matrix associated with time-reversal symmetry.
The above PH transformation results in
\begin{align}
    F_{ij}(q) &= T\sum_n \int \frac{dk}{2\pi}
    [U_T g(-i\omega_n,-k-q/2)^T U_T^{-1}]_{ij} 
    \n 
    &\qquad\qquad \times
    [U_T g(i\omega_n,k-q/2)^T U_T^{-1}]_{ij}.
\end{align}
In the case of a conventional single band superconductor, $g(i\omega_n,k)$ is a c-number and $U_T=1$. Thus we obtain
\begin{align}
    F(q)&= T\sum_n \int \frac{dk}{2\pi}
    g(-i\omega_n,-k-q/2)
    g(i\omega_n,k-q/2) \n
    &=F(-q),
\end{align}
which indicates $q$ is an even function with respect to $q$.
This means the linear coupling term vanishes ($f_1=0$) for a single band superconductor from the PH symmetry~\cite{Tsuji15}, which is the reason why the collective mode has been studied for higher order responses such as THG so far~\cite{Shimano20}.
In contrast, $F(q)$ becomes a matrix in a multiband superconductor, and one can obtain nonvanishing comdponent even under the constraint of PH symmetry.
To see this, let us consider the case of Rice-Mele model which is given by $\xi(k)$ in Eq.~\eqref{eq: xi}.
In this case, the time reversal symmetry is defined with $U_T=1$,
and the PH constraing on $F_{ij}(q)$ reads
\begin{align}
    F_{ij}(q) &= T\sum_n \int \frac{dk}{2\pi}
    g(-i\omega_n,-k-q/2)^T_{ij} 
    g(i\omega_n,k-q/2)^T_{ij} \n
    &=
    F_{ji}(-q).
\end{align}
This indicates that the diagonal component $F_{ii}(q)$ is an even function with respect to $q$, and hence, $f_{1,ii}=0$, while the off-diagonal component $F_{ij}(q)$ is neither even nor odd function of $q$ and $f_{1,ij}\neq 0$ for $i\neq j$ generally.
Moreover, the condition $F_{12}(q)=F_{21}(-q)$ indicates the linear term satisfies $f_{1,12}=-f_{1,21}$. 
Since these terms appear in the effective Lagrangian as a complex conjugate pair, $f_{1,12}$ should be purely imaginary, which is consistent with the result of an explicit computation in Eq.~\eqref{eq: d microscopic}.
According to the discussion in Sec.~\ref{sec: collective}, a purely imaginary $d$ term results in a linear coupling between Leggett mode and an external light field as was observed in Sec.~\ref{sec: application}.  

We note that the $f_0$ term in Eq.~\eqref{eq: F q} gives rise to the coupling term $\epsilon$ in the GL free energy density in Eq.~\eqref{eq: L two band}. Upon the mode expansion, the $\epsilon$ term gives a mass term for the Leggett mode $-\epsilon \Psi_{1,0} \Psi_{2,0} \theta^2$.
Since we obtain nonzero $f_0$ generally, this indicates that the excitation energy of the Leggett mode $E_L$ satisfies $E_L \propto \sqrt{\Psi_{1,0} \Psi_{2,0}}$
once we incorporate the dynamical term ($\partial_t^2 \theta$) in the free energy, which is consistent with the known formula for $E_L$ \cite{Leggett66,Murotani17}.

\section{Symmetry analyses for the collective modes\label{app: symm}}
In this appendix, we discuss the role of the symmetry in the linear coupling between the collective excitations and the electric magnetic field, in terms of the allowed form for $Q_{\mu,\alpha}$ [Eq.~(\ref{eq:q})].

Let us first see how the form of $Q_{\mu,\alpha}$ is constrained
by the inherent particle-hole symmetry. The presence of the particle-hole
symmetry assures the existence of the unitary transformation $U_{\text{PHS}}$
satisfying 
\begin{align}
U_{\text{PHS}}^{\dagger}\mathcal{H}_{\text{BdG}}(k)U_{\text{PHS}}=-\mathcal{H}_{\text{BdG}}^{\ast}(-k).
\end{align}
 In the presence of this symmetry relation, the Green's function are
transformed as 
\begin{align}
U_{\text{PHS}}^{\dagger}G(k,i\omega)U_{\text{PHS}} & =-[G(-k,i\omega)]^{\ast}.
\end{align}
Indeed, for the Rice-Mele model Eq.~(\ref{eq: xi}), the above relation
holds with $U_{\text{PHS}}=\tau_{y,1}+\tau_{y,2}$. Additional relations
$U_{\text{PHS}}^{\dagger}v(k)U_{\text{PHS}}=-[v(-k)]^{\ast}$ and
$U_{\text{PHS}}^{\dagger}\tau_{\mu\alpha}U_{\text{PHS}}=-\tau_{\mu\alpha}^{\ast}$
leads to 
\begin{align}
&Q_{\mu\alpha}(i\Omega) \nonumber\\
& =\int\dfrac{d\omega}{2\pi}\int\dfrac{dk}{(2\pi)^{d}}\text{tr}[v(-k)G(-k,i\omega+i\Omega)\tau_{\mu\alpha}G(-k,i\omega)]^{\ast}\nonumber\\
 & =[Q_{\mu\alpha}(i\Omega)]^{\ast}.
\end{align}
In addition to the particle-hole symmetry, typical superconductors
also have the time-reversal symmetry, which is denoted as
\begin{align}
U_{\text{TRS}}^{\dagger}\mathcal{H}_{\text{BdG}}(k)U_{\text{TRS}} & =\mathcal{H}_{\text{BdG}}^{\ast}(-k),\\
U_{\text{TRS}}^{\dagger}[G(k,i\omega)]U_{\text{TRS}} & =[G(-k,-i\omega)]^{\ast}.
\end{align}
For the Rice-Mele model (\ref{eq: xi}), $U_{\text{TRS}}=1$. Using $v(k)^{\ast}=-v(-k)$,
we obtain 
\begin{align}
Q_{\mu\alpha}(i\Omega) 
& =-\int\dfrac{d\omega}{2\pi}\int\dfrac{dk}{(2\pi)^{d}}\text{tr}[v(-k)G(-k,-i\omega-i\Omega)\nonumber\\&\qquad\qquad\qquad\qquad\qquad\qquad\quad\tau_{\mu\alpha}^{\ast}G(-k,-i\omega)]^{\ast}\nonumber\\
 & =\begin{cases}
-\text{tr}[Q_{\mu\alpha}(-i\Omega)]^{\ast} & \mu=x\\
\text{tr}[Q_{\mu\alpha}(-i\Omega)]^{\ast} & \mu=y
\end{cases}.
\end{align}
Namely, for the amplitude modes $Q_{x\alpha}$, $Q_{x\alpha}(i\Omega)$
must be an odd real function with respective to the Matsubara frequency
(if combined with the PH symmetry), and thus has vanishing coupling
in the static limit $\Omega\to0$ corresponding to the GL description.
While the phase modes $Q_{y\alpha}$ may have a finite value even
in the static limit, the Nambu-Goldstone mode does not appear in physical quantities
as it can be gauged out, so that the multi-gap nature is essential for having nonzero linear response. 

In general, the linear coupling to the Leggett mode can also vanish due
to the presence of an additional symmetry. In the Rice-Mele model,
the model parameters $\delta t$ and $m$ control the presence/absence
of spatial symmetries. Namely, the system becomes site-center inversion
symmetric when $\delta t=0$, while bond-center
inversion symmetic when $m=0$.
While both symmetries relate $k$ with $-k$, the latter symmety accompanies
the interchange of the sublattices.

For the $\delta t=0$ case, the symmetry relation for the Hamiltonian
reads $U_{z}^{\dagger}H(\pi+k)U_{z}=H(\pi-k)$ with 
\begin{align}
U_{z}=\begin{pmatrix}\sigma_{z} & O\\
O & \sigma_{z}
\end{pmatrix}.
\end{align}
Using $U_{z}^{\dagger}v(\pi+k)U_{z}=-v(\pi-k),U_{z}^{\dagger}\tau_{\mu\alpha}U_{z}=\tau_{\mu\alpha}$,
and $U_{z}^{\dagger}G(\pi+k,i\omega)U_{z}=G(\pi-k,i\omega)$, we obtain
\begin{align}
Q_{\mu\alpha}(i\Omega) 
& =-\int\dfrac{d\omega}{2\pi}\int\dfrac{dk}{2\pi}\text{tr}[v(2\pi-k)\nonumber\\
&\qquad\qquad\quad G(2\pi-k,i\omega+i\Omega)\tau_{\mu\alpha}G(2\pi-k),i\omega)]\nonumber\\
 & =-Q_{\mu\alpha}(i\Omega),
\end{align}
i.e., the linear coupling must identically vanish for any collective
modes. On the other hand, for the $m=0$ case, the symmetry relation
reads $U_{x}^{\dagger}H(k)U_{x}=H(-k)$ with 
\begin{align}
U_{x}=\begin{pmatrix}\sigma_{x} & O\\
O & \sigma_{x}
\end{pmatrix}.
\end{align}
 Since this unitary interchanges the sublattice, we have $U_{x}^{\dagger}\tau_{\mu1}U_{x}=\tau_{\mu2}$
and $U_{x}^{\dagger}\tau_{\mu2}U_{x}=\tau_{\mu1}$. The transformation
for other matrices are the same as the above case, i.e., $U_{x}^{\dagger}v(k)U_{x}=-v(-k)$
and $U_{x}^{\dagger}G(k,i\omega)U_{x}=G(-k,i\omega)$. We obtain
\begin{align}
&Q_{\mu1}(i\Omega)  \nonumber\\& =-\int\dfrac{d\omega}{2\pi}\int\dfrac{dk}{2\pi}\text{tr}[v(-k)G(-k,i\omega+i\Omega)\tau_{\mu2}G(-k,i\omega)]\nonumber\\
 & =-Q_{\mu2}(i\Omega).
\end{align}
This relation implies that the oscillation of the relative phase (or amplitude) between $\Delta_1$ and $\Delta_2$ is allowed to couple with the electromagnetic field.

\bibliography{higgs}

\end{document}